\documentclass[12pt,tightenlines,eqsecnum,floats,aps,amsmath,amssymb,nofootinbib,prd]{revtex4}

\usepackage{setspace}
\usepackage{amsmath,amssymb,amsfonts,amsthm}
\usepackage{graphicx}
\usepackage{enumerate}

\usepackage[pdftex]{hyperref}

\def\be{\begin{equation}}
\def\ee{\end{equation}}
\def\ba{\begin{eqnarray}}
\def\ea{\end{eqnarray}}
\def\bi{\begin{itemize}}
\def\ei{\end{itemize}}

\def\qo{\mathring{q}}
\def\Do{\mathring{D}}
\def\Eo{\mathring{E}}
\def\xh{\boldsymbol{n}}
\def\e{\epsilon}

\def\epso{\mathring{\epsilon}}

\def\R{\mathcal{R}}
\def\N{N}

\def\D{\mathcal{D}}
\def\L{\mathcal{L}}
\def\LT{\Lambda_{\vec{N}}}
\def\LN{\Lambda_{R}}
\def\LB{\Lambda_{B}}
\def\xh{\hat{x}}
\def\t{\tau}
\def\odd{{\rm odd}}
\def\even{{\rm even}}
\def\LNo{\mathring{\Lambda}_{R}}
\def\Ei{g}
\def\Eii{h}
\def\Dadm{D_{\text{\tiny ADM}}}
\def\lN{\lambda_{\vec{N}}}
\def\lR{\lambda_{\vec{R}}}

\def\Lnu{\Lambda_{\vec{\nu}}}

\def\LBo{\mathring{\Lambda}_{B}}
\def\Hadm{H_{\text{\tiny ADM}}}
\def\Nadm{N_{\text{\tiny ADM}}}
\def\Hadmo{H^0_{\text{\tiny ADM}}}

\begin{document}

\title{Note on the  phase space of asymptotically flat gravity in Ashtekar-Barbero variables}

\author{Miguel Campiglia} \email{campi@fisica.edu.uy}
 \affiliation{Raman Research Institute \\ Bangalore 560080, India}
 \affiliation{Instituto de F\'isica, Facultad de Ciencias \\  Montevideo 11400, Uruguay}

% \affiliation{Raman Research Institute \\Bangalore-560 080, India}

\begin{abstract}

We describe the canonical phase space of  asymptotically flat gravity  in Ashtekar-Barbero variables. We show that the Gauss constraint multiplier must fall off slower than previously considered in order to recover ADM phase space. The  generators of the asymptotic Poincare group are derived within the Ashtekar-Barbero phase space without reference to the ADM generators. The resulting expressions are shown to agree, modulo Gauss constraint terms,  with those obtained from the ADM generators. A  payoff of this procedure  is  a new expression for the generator of asymptotic rotations which is polynomial in the triad and hence better suited for quantum theory.  Our  treatment  complements  earlier   description  by Thiemann in the context of self-dual variables.

\end{abstract}
\maketitle

\section{Introduction}
In \cite{Taf} Thiemann adapted the  description of asymptotically flat canonical gravity in ADM variables \cite{adm} given in  \cite{RT,BO} to Ashtekar variables \cite{newvar,Aprd87}.  Among other things, he obtained the  generators of the asymptotic Poincare group, showed their agreement with the ADM generators, and verified their Poisson brackets reproduce the Poincare algebra.

Here we revisit Thiemann's analysis in the context of real $SU(2)$ variables \cite{barbero}  with an arbitrary Barbero-Immirzi parameter $\gamma$ \cite{immirzi,ss}.   %The treatment however applies to the $\gamma=i$ case, which is how we  compare our expressions with those in \cite{Taf}. % As we shall see, no subtleties arise from the additional `Lorentzian'  term in the Hamiltonian constraint that is present when  $\gamma \neq \pm i$. 
We take the same parity conditions and leading fall-off terms  as those used in \cite{Taf,RT}. For subleading terms we use the  more general fall-offs given in \cite{BO}. We also show that, independently of the choice of subleading fall-offs, the leading term of the Gauss constraint multiplier must  fall off slower than what was considered in \cite{Taf} if one wants to recover ADM phase space. 

To obtain the Poincare generators, we follow the strategy of \cite{RT} where one seeks for boundary terms to the Hamiltonian and diffeomorphism constraint that yield well defined phase space functions when then lapse and shift have asymptotic values that correspond to Poincare transformations. This strategy is followed within the Ashtekar-Barbero phase space, without resorting to the ADM expressions.

For spacetime translations, we recover the known expressions given in  \cite{Aprd87,Taf}. For boosts  we obtain a  generator,  Eq. (\ref{finalC}), that is shown to agree on the Gauss constraint surface with the one obtained in \cite{Taf}. The situation is the most subtle with rotations. In \cite{Taf}  the generator of rotations was obtained from the ADM generator. The resulting expression involves the spin connection and hence is non-polynomial in the triad.  On the other hand the generator obtained here, Eq. (\ref{finalD}),  is polynomial in the triad.  Showing that the two agree (modulo a Gauss constraint term with phase-space dependent multiplier) requires  careful  comparison of the expressions. %Finally, for boosts we obtain an  expression (Eq. (\ref{finalC})) whose  equivalence with   the generator given in \cite{Taf} is rather straightforward to establish.  

The motivation for the present study comes from its application to quantum theory.  In particular, the expression for  angular momentum  obtained here facilitates  the unitary implementation of asymptotic rotations described  in \cite{mm3}.

The organization of the material is as follows.  In section \ref{sec2} we  review  the asymptotically flat ADM phase space as treated in \cite{BO}.  The section serves to set up notation, display the ADM  Poincare generators for later comparison, and present the guiding principle of references \cite{RT,BO} that we  follow in section \ref{sec3}.  In section \ref{sec3A} we describe the  Ashtekar-Barbero  phase space counterpart of the ADM phase space of section \ref{sec2}. In \ref{sec3B1} we discuss the Gauss constraint and corresponding asymptotic behavior of its multiplier. In \ref{sec3B2} we discuss the Hamiltonian and diffeomorphism constraints, and in \ref{sec3C} the Poincare generators. The discussion of  rotations will be more detailed than for the other generators, since it is here that the comparison with ADM and Thiemann's expression is more subtle. In section \ref{sec4} we show the Poincare generators of section \ref{sec3} coincide with the ADM and Thiemann ones. %Some calculations are given in the appendices. 

\section{Review of  ADM case} \label{sec2}
\subsubsection{Phase space} \label{ADMps} \label{sec2A}
In the asymptotically flat case, the  Cauchy slice  $\Sigma$ is such that it admits, outside a compact set, cartesian coordinates $x^I$, $I=1,2,3$ that extend to infinity. Let $\qo_{ab}$ be the flat background metric associated with the cartesian coordinates so that  $\qo_{IJ} = \delta_{IJ}$. Let  $r:= \sqrt{(x^1)^2+(x^2)^2+(x^3)^2}$ and $\xh^I:= x^I/r$. The phase space is then given by the standard canonical pair $(q_{ab}, \pi^{ab})$  satisfying the following fall-off conditions in the cartesian coordinate system:\footnote{$\Sigma$, $q_{ab}$ and $\pi^{ab}$ are taken to be $C^{\infty}$.   A tensor $f$ is  $O(r^{-\beta})$ (denoted by $O^{\infty}(r^{-\beta})$ in \cite{BO}) if for $n=0,1,\ldots$, the $n$-th partial derivatives of $f$ in the Cartesian chart, $\partial_{I_n} \ldots \partial_{I_1}f$, are bounded by $c_n r^{-n-\beta}$ for constants $c_n$.}
\ba
q_{IJ} & = &\qo_{IJ} +\frac{h_{IJ}(\xh)}{r}+O(r^{-1-\e}) \label{falloffq},\\
\pi^{IJ} & = &\frac{p^{IJ}(\xh)}{r^2}+O(r^{-2-\e})\label{falloffpi},
\ea
where $\e>0$ and  $h_{IJ}$ and $p^{IJ}$ are of even and odd parity respectively:
\ba
 h_{IJ}(-\xh) &= & h_{IJ}(\xh) \label{parityq},\\
 p^{IJ}(-\xh) &= & -p^{IJ}(\xh) . \label{paritypi}
\ea
%The justification for the particular fall-off and parity conditions is given at the end of the next subsection.
%At the end of next section we will argue these are the  `minimal' conditions such that the symplectic structure is well defined (non-divergent) and which allows for well defined and non-trivial Poincare generators/charges.

The fall-off conditions ensure the symplectic structure,
\be
\Omega(\delta_1,\delta_2)=\frac{1}{16 \pi G} \int_{\Sigma}(\delta_1 q_{ab}\delta_2 \pi^{ab}-\delta_2 q_{ab}\delta_1 \pi^{ab}), \label{sympf}
\ee
is well defined, and allows the existence of non-trivial Poincare generators \cite{RT,BO}. 

We will be dealing with phase space functions $F[q,\pi]$ that are integrals over $\Sigma$ of local functions of $q_{ab}$ and $\pi^{ab}$.  The two basic conditions that are required for such expressions are:
\begin{itemize}
\item[(i)]  $F$ should be finite, i.e., the integral over $\Sigma$ should be convergent
\item[(ii)] $F$ should admit a Hamiltonian vector field, i.e. $\delta F= \Omega(\delta_F,\delta) \; \forall \delta$ .
\end{itemize}
Above $\delta = (\delta q_{ab},\delta \pi^{ab})$ is any variation and $\delta_F = (\delta_F q_{ab},\delta_F \pi^{ab})$ is the Hamiltonian vector field of $F$. Both $\delta$ and $\delta_F$ are vector fields in phase space and hence respect the  fall-off and parity conditions given above.  Condition (ii) encompasses the `functional differentiability' requirement that  $\delta F$ contains no surface terms, and the requirement that  the action of $F$ preserves the fall-off and parity conditions. Finally, given two functions $F$ and $G$ satisfying (i) and (ii), their Poisson bracket is defined by $\{F,G\} := \Omega(\delta_F,\delta_G)$.

\subsubsection{Constraints and Poincare generators} \label{pgadm} \label{sec2B}
In \cite{BO} it is shown that the Hamiltonian and diffeomorphism constraints:
\ba
H_0[N] & := & \frac{1}{16 \pi G} \int_{\Sigma} N (q^{-1/2}(\pi_{ab}\pi^{ab}-\tfrac{1}{2}\pi^2)-q^{1/2} \R ) \label{H0}, \\
D_0[\vec{N}] &:=&-\frac{1}{8 \pi G} \int_{\Sigma} N_a D_b \pi^{ab} \label{D0},
\ea
satisfy (i) and (ii) when  the lapse and shift have the following $r \to \infty$ asymptotic behavior: 
\ba
N & = & S(\xh)+O(r^{-\e}) ,\label{fallofflapse}\\
N^I & = & S^I(\xh)+O(r^{-\e}) ,\label{falloffshift}
\ea
with
\ba
S(-\xh) & = & -S(\xh), \\
S^I(-\xh) & = & -S^I(\xh). \label{oddst}
\ea
 $H_0[N]$ and $D_0[\vec{N}]$ with lapse and shift obeying (\ref{fallofflapse}), (\ref{falloffshift}) generate the gauge transformation of the theory.

On the other hand,  asymptotic Poincare transformations correspond to  lapse and shift satisfying the asymptotic conditions:
\ba
N & \to & \alpha + B  \label{lapseP}+ \ldots \label{asymlapse}\\
N^I &\to & \alpha^I+ R^I\label{shiftP} +\ldots , \label{asymshift}
\ea
where $\alpha$, $\alpha^I$, are constants that represents spacetime translations, 
\ba
B & =&\beta_I x^I   \\
R^I &=&\beta^I_{\phantom{I} J} x^J \label{Rbeta}
\ea
with constant $\beta$'s  and  $\beta_{JI}=-\beta_{IJ}$ represent boost and rotations, and the dots indicates `pure gauge' terms  as in (\ref{fallofflapse}), (\ref{falloffshift}).

In \cite{BO} it is shown that the generators of the asymptotic Poincare transformations  satisfying (i) and (ii) above are given by:
\ba
H[N]&:=&H_0[N]+ \frac{1}{8 \pi G}\oint_{\infty} d S_d q^{1/2} q^{a[b}q^{c]d}(N \Do_b q_{c a}-\Do_b N( q_{c a}-\qo_{ca})) \label{Htotal},\\
D[\vec{N}]&:=&D_0[\vec{N}]+\frac{1}{8 \pi G}\oint_{\infty} d S_a N_b \pi^{ab} \label{Dtotal} =  \frac{1}{16 \pi G} \int_{\Sigma} \pi^{ab}\L_{\vec{N}} q_{ab}, 
\ea
where $\oint_{\infty} \equiv \lim_{r \to \infty} \oint_{S_r}$ with $S_r$ the 2-sphere of radius $r$ with respect to the cartesian system $x^I$ and $\Do$ the derivative compatible with $\qo_{ab}$.

We emphasize that it is the `total' $H[N]$ and  $D[\vec{N}]$ that satisfy (i) and (ii).  The writing of the generators as `volume  plus surface term'  is for convenience;  each term is not in itself a well defined phase space function.  In particular,  the surface integrals can be divergent.\footnote{This can happen for asymptotic rotations and boosts, and for phase space points outside the constraint surface $H_0 = D_0=0$. This `off-shell' divergence of surface terms does not occur with the fall-off conditions used in \cite{RT}, which are schematically of the form $q_{ab} = \qo_{ab} +(\even) r^{-1}+O(r^{-2})+O(r^{-2-\e})$ and $\pi^{ab}= (\odd) r^{-2}+O(r^{-3})+O(r^{-3-\e})$.} % These more restrictive conditions have the disadvantage of not representing phase space as a cotangent bundle, and may also be less suited for quantum theory as discussed in section 5E of \cite{mm3}.}  
On the constraint surface however, finiteness of the generators imply finiteness of the surface integrals, which give then the value of the corresponding Poincare charges (e.g. angular momentum in the case of rotations).

%The values of the generators on the constraint surface $H_0 \approx 0, D_0 \approx 0$ are given by the surface  terms in (\ref{Htotal}) and (\ref{Dtotal}). With these expressions one can justify the choice of fall-off and parity conditions as follows. First, in order to allow for non-trivial energy, we need the metric deviation to fall off as $r^{-1}$ (think for instance of the $t=$ constant slices of Schwarzschild metric). Second, to allow for non-trivial linear momentum, we need $\pi^{IJ}$ to fall off as $r^{-2}$. But this implies a divergent $r^{-3}$ fall-off in the symplectic structure (\ref{sympf}).  This $r^{-3}$ divergent contribution is then eliminated by the parity conditions (\ref{parityq}),(\ref{paritypi})  (the reverse parity choice is not satisfactory since it would lead to  zero linear momentum and energy).

% \emph{Side comment:} The treatment of asymptotically AdS case is goes parallel with the present one, just with different falloff behavior of the fields (which do not require the use of odd/even functions!). In particular the boundary terms have exactly the same form as in (\ref{Htotal}),(\ref{Dtotal}) with $\qo$  the background hyperbolic 3-metric (corresponding to a $t= $ constant slice of AdS spacetime) and with asymptotic behavior of lapse and shift associated with the AdS isometries. 

\section{Ashtekar-Barbero variables} \label{sec3}
%Already in the original Ashtekar's articles there is discussion on the asymptotically flat case  \cite{Aprd87} . However only asymptotic translations (spatial and temporal) are contemplated. This was amended by Thiemann \cite{Taf} by incorporating asymptotic Lorentz transformation in the formulation.  The way Thiemann proceedes is roughly to rewrite the  ADM expressions (\ref{Htotal}, \ref{Dtotal}) in terms of connection variables.  We will instead proceed along the lines of \cite{RT} of starting with the standard constraints and then finding suitable modification that renders it well defined.

\subsection{Phase space} \label{sec3A}
The Cauchy slice $\Sigma$ and the cartesian coordinates $\{x^I\}$ are taken as in the previous section.  The canonical pair is now given by an $su(2)$ connection one-form $A_a=A_a^i \t_i$ and conjugate electric field $E^a=E^a_i \t_i$, with $\t_i,\; i=1,2,3$ an $su(2)$ basis satisfying $[\t_i,\t_j]=\epsilon_{i j k} \t_k$. As  will shortly become clear, in order to have a well defined symplectic structure it is necessary that the electric fields asymptote to a  fixed densitized triad $\Eo^a_i$ (whose associated metric is taken to agree with  $\qo_{ab}$ of Eq. (\ref{falloffq})). We chose this fixed, zeroth order asymptotic electric field to be given by  $\Eo^I_i=\delta^I_i$.  The phase space is  then given by pairs $(A_a^i,E^a_i)$  satisfying the analogue of the ADM asymptotic conditions \cite{Taf}:
 \ba
E^I_i & = &\Eo^I_i+\frac{f^I_i(\xh)}{r}+O(r^{-1-\e}) \label{falloffE},\\
A_I^i & = &\frac{a_{I}^i(\xh)}{r^2}+O(r^{-2-\e})\label{falloffA},
\ea
with
\ba
f^I_i(-\xh) &= & f^I_i(\xh) \\
a_{I}^i(-\xh) &= & -a_{I}^i(\xh) .
\ea
It is not difficult to verify that these fall-offs and parity conditions imply the ADM ones described in the previous section. The fall-off conditions ensure  the symplectic structure, % (see Appendix \ref{symstrapp})
\be
\Omega(\delta_1,\delta_2):=\frac{1}{8\pi G \gamma} \int_{\Sigma}(\delta_1 A_a^i \delta_2 E^{a}_i-\delta_2 A_a^i \delta_1 E^{a}_i) ,\label{sympfnv}
\ee
 is well defined. We now see the need to keep fixed the zeroth order electric field in (\ref{falloffE}): Had we allowed for all possible $SU(2)$-rotated $\Eo^a_i$'s (so that the asymptotic metric still satisfies (\ref{falloffq})), we could not have ensured convergence of the integral (\ref{sympfnv}).

Below we will be dealing with phase space functions $F[A,E]$ that are integrals over $\Sigma$ of local functions of $A_a^i$ and $E^a_i$. Such functions will be required to satisfy the conditions (i) and (ii) described at the end of section \ref{sec2A} (now with respect to the sympectic form (\ref{sympfnv})).

\subsection{Constraints}\label{sec3B1}
\subsubsection{Gauss Constraint} \label{gausssec} \label{sec3B1}
In connection variables, there appears the additional Gauss law constraint
\be
G[\Lambda]:=\frac{1}{8 \pi G \gamma}\int_\Sigma \Lambda^i \D_a E^a_i ,\label{gauss}
\ee
where $\D_a$ is the covariant derivative associated to the connection $A_a^i$, acting as $\D_a f_i  = \partial_a f_i+\e_{i j k}A_a^j f^k$. 
Since both terms in  $\D_a E^a_i$ fall off as $(\odd) r^{-2}+O(r^{-2-\e})$, the minimal condition on the multiplier $\Lambda^i$ ensuring convergence of the integral is:
\be
\Lambda^i = \frac{\lambda^i(\xh)}{r}+O(r^{-1-\e}) \label{fallofflambda}
\ee
with
\be
\lambda^i(-\xh)=\lambda^i(\xh). \label{paritylambda}
\ee
We now verify that with this fall-off and parity condition,  $G[\Lambda]$ also satisfies (ii):
\ba
\delta G & = & \frac{1}{8 \pi G \gamma}\int_{\Sigma} \Lambda^i( \partial_a \delta E^a_i +\e_{i j k}A_a^j \delta E^a_k+\e_{i j k}\delta A_a^j E^a_k) \label{G1} \\
&=&  \frac{1}{8 \pi G \gamma}\int_{\Sigma} (\delta_\Lambda A_a^i \delta E^a_i -\delta A_a^i \delta_\Lambda E^a_i) \equiv \Omega(\delta_\Lambda,\delta) \label{G2}
\ea
where
\ba
\delta_\Lambda A_a^i & =& -\partial_a \Lambda^i +\e_{ijk} \Lambda^j A_a^k \label{deltaLA}=-\D_a \Lambda^i \label{deltaLA}\\
\delta_\Lambda E^a_i & =& \e_{i j k} \Lambda^j  E^a_k. \label{deltaLE}
\ea
In going from (\ref{G1}) to (\ref{G2}), we performed the integration by parts:
\be
\int_{\Sigma} \Lambda^i \partial_a \delta E^a_i   = \oint_\infty dS_a \Lambda^i \delta E^a_i - \int_{\Sigma} \partial_a \Lambda^i \delta E^a_i =- \int_{\Sigma} \partial_a \Lambda^i \delta E^a_i
\ee
where the surface term being $(\odd) (\even)r^{-1} (\even)r^{-1}+ O(r^{-2-\e})$ vanishes. It is easy to verify that (\ref{deltaLA}) and (\ref{deltaLE}) preserve the fall-off and parity conditions, and hence is a  well defined phase space variation.  Finally, the relation $\{G[\Lambda],G[\Lambda'] \} = G[[\Lambda,\Lambda']]$ can be verified thanks to the vanishing of the  surface term:
\be
\oint_{\infty}dS_a \epsilon_{ijk} E^a_i \Lambda_j \Lambda'_k =0,
\ee
as implied by the fall-off and parity condition (\ref{fallofflambda}).

We now show that the leading term in  (\ref{fallofflambda})  is crucial for the recovery of ADM phase space in that it accounts for `pure $SU(2)$ gauge'  components of the $r^{-1}$ term of the triad (\ref{falloffE}).  The  doubly densitized inverse metric is given by $\tilde{\tilde{q}}^{IJ} = E^I_i E^J_i = \Eo^I_i \Eo^J_i+ 2 \Eo^{(I}_i f^{J)}_i/r +O(r^{-1-\e})$  
 from which it follows that
\be 
q_{IJ} =  \Eo^I_i \Eo^J_i- 2 \Eo^{(I}_i f^{J)}_i/r +O(r^{-1-\e}). \label{qE}
\ee
Define $f_{IJ}:=-2 \Eo^{I}_i f^{J}_i \equiv - 2 f^{J}_I$. Equating (\ref{qE}) with (\ref{falloffq}) we conclude that the $r^{-1}$ term of the metric is given by: $h_{IJ}=f_{(IJ)}$.  From this perspective $f_{[IJ]}$ appear as `pure gauge' components.

On the other hand, the variation of $f_{IJ}$ under $SU(2)$ gauge transformation can be found  by substituting (\ref{falloffE}) and (\ref{fallofflambda}) in (\ref{deltaLE}): 
\be
\delta_\Lambda f^I_i = \epsilon_{i j k} \lambda^j \Eo^{I}_k \implies \delta_\Lambda f_{IJ}= 2 \e_{IJ k} \lambda^k ,
\ee
which is in agreement with the previous `pure gauge' interpretation of $f_{[IJ]}$.

In the following sections we will often encounter Gauss constraints (\ref{gauss}) smeared with phase space dependent multipliers. We now verify properties (i) and (ii) are still satisfied in such cases. For $\Lambda^i=\Lambda^i(A,E)$ satisfying (\ref{fallofflambda}) and (\ref{paritylambda}), finiteness follows by the some fall-off/parity argument given before. From (\ref{G2}) it follows that the variation of $G[\Lambda]$ is now given by

\be
\delta G[\Lambda] = \Omega(\delta_{\Lambda(A,E)},\delta) + \frac{1}{8 \pi G \gamma}\int_\Sigma \delta \Lambda^i(A,E) \D_a E^a_i . \label{deltaGgral}
\ee
If $\Lambda^i=\Lambda^i(A,E)$ does not depend on derivatives of $A^i_a$ or $E^a_i$, then (\ref{deltaGgral}) is already differentiable. Otherwise  one needs to integrate by parts the second term in (\ref{deltaGgral}) to obtain an integrand that does not depend on derivatives of the variations $\delta A_a^i$, $\delta E^a_i$. We now argue that the corresponding surface terms will always be $(\odd) r^{-2}+ O(r^{-2-\e})$ and hence vanish. Consider a term in $\delta \Lambda^i$ of the form $F_{a i j}^b[A,E] \partial_b \delta E^{a}_j$.  Since $\delta \Lambda^i=(\even) r^{-1}$ and $ \partial_b \delta E^{a}_j = (\odd) r^{-2}$ it follows that $F_{a i j}^b=(\odd) r$. The surface term will then be $d S_b F_{a i j}^b \delta E^{a}_j \D_c E^c_i= (\odd) r^{-2}+O(r^{-2-\e})$. A similar argument shows that the surface term coming from a variation involving a derivative of $\delta A_a^i$ also vanishes. The  argument may also be extended to allow for  derivatives of higher order, but the above considerations are  enough for our purposes since all phase space dependent multipliers we encounter depend at most on  first derivatives of the canonical variables. Thus, we conclude that the Gauss constraint with phase space dependent multiplier satisfying (\ref{fallofflambda}) and (\ref{paritylambda}) is  differentiable. Finally, it is easy to verify from the above expressions that the contribution to the Hamiltonian vector field coming  from the second term in (\ref{deltaGgral})  also preserves the fall-off and parity conditions.

As a final note, we point out that the alternative expression of the Gauss constraint obtained by integration by parts in (\ref{gauss}) leads to
\be
G[\Lambda]=-\frac{1}{8 \pi G \gamma}\int_\Sigma E^a_i  \D_a\Lambda^i  +\frac{1}{8 \pi G \gamma} \oint_{\infty} dS_a  E^a_i \Lambda_i. \label{gaussibp}
\ee
Whereas the full $G[\Lambda]$ is well defined, the two terms in the RHS (\ref{gaussibp}) are not necessarily well defined by themselves. Indeed, it can be easily seen that the fall-offs (\ref{fallofflambda}) do not ensure convergence of the surface term in (\ref{gaussibp}).
\subsubsection{Hamiltonian and diffeomorphism constraints} \label{sec3B2}
We start with the following form of the Hamiltonian and diffeomorphism constraints \cite{alrev}:
\ba
H_0[\N] & := & \frac{1}{16 \pi G} \int_{\Sigma}  \N  (\e_{i j k} F_{ab}^i E^a_j E^b_k -2(1+\gamma^2)K^i_{[a}K^j_{b]}E^a_i E^b_j) \label{H0nv}, \\
D_0[\vec{N}] &: = &\frac{1}{8 \pi G \gamma} \int_{\Sigma} E^a_i \L_{\vec{N}} A_a^i\label{D0nv}, 
\ea
where $F^i_{ab}:= \partial_a A^i_b-\partial_b A_a^i + \epsilon_{ijk}A^j_a A^k_b$, $K_a^i:=\gamma^{-1}(A_a^i-\Gamma_a^i)$ and $N$ of  density weight $-1$.  The relation between  the constraints $H_0$ and $D_0$ of this section and those of section \ref{sec2} will be  described in section \ref{sec4}.

The minimal conditions for (\ref{H0nv}) and (\ref{D0nv}) to be finite are as in the ADM case:
\ba
N & = & S(\xh)+O(r^{-\e}) ,\label{folapse}\\
N^I & = & S^I(\xh)+O(r^{-\e}) , \label{foshift}
\ea
with $S$ and $S^I$ odd. It is  easy to verify that  $D_0$ and the first term in $H_0$  satisfy (ii). We now argue the second, `KKEE' term in $H_0$  also satisfies (ii). Under  variations of this term,  the potentially problematic surface contribution come from derivatives of the triad in $\Gamma_a^i$. Schematically:
\be
(\int N E E K \delta \Gamma)_{\text{Bdy}}  = \oint N E E K \delta E =0, \label{bdylor}
\ee
where the vanishing occurs since the integrand of the surface term  falls off as $r^{-3}$. Finally, it is easy to verify that the contribution from the KKEE piece to the Hamiltonian vector field preserves the fall off and parity conditions.

$G[\Lambda]$, $H_0[\N]$ and $D_0[\vec{N}]$ with multipliers satisfying the conditions above are the constraints/gauge generators of the theory.

\subsection{Poincare generators} \label{sec3C}
We now want to extend $H_0$ and $D_0$ in order to obtain well defined generators for  lapse and shift corresponding to asymptotic Poincare transformations:
\ba
N & \to & \alpha + B  +\ldots \label{lP}\\
N^I &\to & \alpha^I+ R^I + \ldots \label{sP},
\ea
with $B=\beta_I x^I$, $R^I =\beta^I_{\phantom{I} J} x^J$ as in section \ref{sec2B} and the dots indicate gauge terms (\ref{folapse}), (\ref{foshift}). Following the strategy of \cite{RT} we will start by adding surface terms that cancel the unwanted boundary contribution of the variations of $H_0$ and $D_0$.

First, we notice that the  `KKEE' term of the Hamiltonian constraint (\ref{H0nv}) is still well defined for the  more general lapse (\ref{lP}): The leading term in the lapse is now $(\odd) r$ so that ${\rm NKKEE }\sim (\odd) r^{-3}+O(r^{-3-\e})$ and the integral converges; the potentially problematic surface term (\ref{bdylor}) is now $(\odd) r^{-2}+O(r^{-2 -\e})$ and  again vanishes. It is also easy to verify that the corresponding Hamiltonian vector field preserves the fall off and parity conditions. 

Thus, the surface terms that cancel the unwanted boundary contributions are the same as in the self-dual formulation \cite{Aprd87}: 
\ba
H_1[\N] & :=& H_0[\N] - \frac{1}{8 \pi G}\oint_{\infty} d S_a \N \e_{i j k}A^i_b E^{a}_j E^{b}_k , \label{H1}, \\
D_1[\vec{N}] & := &D_0[\vec{N}]- \frac{1}{8 \pi G \gamma }\oint_{\infty} d S_a N^a A^i_b E^{b}_i =  -\frac{1}{8 \pi G \gamma} \int_{\Sigma} A^i_a \L_{\vec{N}}E^a_i \label{D1} .
\ea
For the case of asymptotic spacetime translations (so that $B=0$ and $R^I=0$ in (\ref{lP}) and (\ref{sP})), $H_1$ and $D_1$ yield well defined phase space generators which agree with the ADM ones \cite{Aprd87,Taf}. This result will be recovered as a particular case of the general Poincare generators discussed below.

At a formal level, even for boosts and rotations the variations of $H_1$ and $D_1$ have no surface terms. However $H_1$ and $D_1$  are no longer guaranteed to be finite. 

% However, one finds the expressions  are divergent in the case of asymptotic rotations or boosts. Thus $C_1$ and $V_1$  can only be taken as the generators of asymptotic translations, as in \cite{Aprd87}.  

As pointed out in \cite{Taf}, the reason these functions are not well defined for nonzero rotations or boosts becomes clear when one realizes their action would change the zeroth order part of the triad and thus map us out of  phase space. This suggests one should modify the expression by adding a suitable Gauss piece in such a way that the zeroth order part of the triad is kept fixed. In the following we implement this idea.

%The reason these functions are not well defined for nonzero rotations or boosts becomes clearer when we realize their action would change the zeroth order part of the triad and thus map us out of the phase space: Unlike the zeroth order metric, the zeroth order triad is not invariant under an asymptotic rotation or boost. 

\subsubsection{Rotations}
It will be convenient to work with the last expression in (\ref{D1}). When the shift has a nonzero rotation at infinity, the integrand has the following asymptotic behavior:
\be
-A_I^i \L_{\vec{N}}E^I_i = A_I^i E^J_i \partial_J N^I+ \ldots   \; \xrightarrow[r \to \infty]{}  \, A_I^i  \Eo^J_i \beta^I_{\phantom{I} J} +  (\odd) r^{-3} +O(r^{-3-\e}) .\label{lieNE}
\ee
The first term falls off as $(\odd)r^{-2}+O(r^{-2-\e})$ and thus we cannot  ensure converge of the integral. This is the same term responsible for rotating the zeroth order part of the triad. In order to compensate, we subtract an appropriate Gauss term $G[\LN]$ with  
\be
\LN^i = \LNo^i +\Lambda^i, \label{fallLN}
\ee
where
\be
\LNo^i:=  -\frac{1}{2}\e_{i j k} \Eo^j_I \Eo^J_k \beta^I_{\phantom{I} J} =\frac{1}{2} \e_{ijk}\Eo^j_a \L_{\vec{R}} \Eo^a_k \label{lambdashift}
\ee
is a constant ($\partial_a \LNo^i=0$), zeroth order term, and  $\Lambda^i$  a `pure gauge' multiplier as in (\ref{fallofflambda}), (\ref{paritylambda}).

By subtracting   $G[\LN]$ from $D_1[\vec{N}]$ we cancel the term responsible for the divergence in (\ref{lieNE}). This also introduces a new divergent and non-differentiable piece which is removed by including  an appropriate boundary term. The final  expression is: %\footnote{It should be noted that $D_1[\vec{N}]$ and  $G[\LN]$ are not well defined individually (unless $R^I=0$). The first two terms in  (\ref{finalD}) is a shorthand notation for the total integral (\ref{D1mG}). If $R^I=0$, $G[\LN]$ becomes a proper Gauss term, the surface term vanishes, and one recovers (\ref{D1}) (modulo a possible `pure gauge' Gauss term). \label{fnote2}}
\ba
D[\vec{N}] &:=&  D_1[\vec{N}]- G[\LN]+\frac{1}{8 \pi G \gamma }\oint_{\infty} d S_a E^a_i \LNo^i . \label{finalD} 
\ea
We now verify (\ref{finalD}) satisfies (i) and (ii).  In section \ref{sec4} we show (\ref{finalD}) agrees, modulo pure gauge Gauss constraint terms, with ADM and Thiemann's expressions. 

To show finiteness, we write (\ref{finalD}) as a volume integral:
\be
D[\vec{N}] =\frac{1}{8 \pi G \gamma} \int_\Sigma \left(- A^i_a \L_{\vec{N}}E^a_i - \e_{ijk} \LN^i  A^j_a E^a_k-(\LN^i-\LNo^i) \partial_a E^i \right),\label{D1mG}
\ee
where we used that $\partial_a (\LNo^i E^a_i)=\LNo^i \partial_a E^a_i$. By construction the first two terms in (\ref{D1mG}) combine to give a convergent fall-off:
\be
- A^i_I \L_{\vec{N}}E^I_i - \e_{ijk} \LN^i  A^j_I E^I_k   = (\odd) r^{-3} +O(r^{-3-\e}) ,\label{asymD}
\ee
where the cancelation of the would-be divergent terms can be explicitly verified by substituting (\ref{lambdashift}) in (\ref{asymD}). The last term in (\ref{D1mG}) is clearly convergent.
We thus conclude (\ref{finalD}) is finite. Let us now verify (ii). As  mentioned earlier  the first term in (\ref{D1mG}) is functional differentiable.  By the same arguments given for the differentiability of the Gauss constraint one finds that the last term in (\ref{D1mG}) is also functional differentiable. The total variation can finally be written as 
\be
\delta D[\vec{N}] = \Omega(\delta_{D[\vec{N}]},\delta),
\ee
with
\ba
\delta_{D[\vec{N}]} A_a^i & = & \L_{\vec{N}}A_a^i -\delta_{\LN} A_a^i  \label{deltaDA}\\
\delta_{D[\vec{N}]} E^a_i & = & \L_{\vec{N}}E^a_i-\delta_{\LN} E^a_i ,\label{deltaDE}
\ea
where  $\delta_{\LN}$ is given by  (\ref{deltaLA}), (\ref{deltaLE}) with $\Lambda=\LN$. It is easy to verify that $\delta_{D[\vec{N}]}$ preserves the falloff and parity conditions.\footnote{Each term in (\ref{deltaDA}) is a well defined variation. For (\ref{deltaDE}) only the total expression is a valid variation,  but not each term independently (except when $R^I=0$).}

Note that in the  above discussion the shift was of the general type (\ref{sP}). If  $R^I=0$ then $\LNo^i=0$ and we recover the generator of translations (\ref{D1}) (up to a possible `pure gauge' Gauss term). 

We conclude the section by verifying  (\ref{finalD}) is $SU(2)$ gauge invariant in the sense that it weakly commutes with the Gauss constraint. Using  Eqns. (\ref{deltaLA}), (\ref{deltaLE}), (\ref{deltaDA}), (\ref{deltaDE}), one finds:
\be
\{ D[\vec{N}] , G[\Lambda]\}= G[\L_{\vec{N}} \Lambda -[\LN, \Lambda]] - \frac{1}{8 \pi G \gamma }\oint_{\infty} d S_a E^a_i (\L_{\vec{N}} \Lambda^i -[\LN, \Lambda]^i). \label{DG}
\ee
The multiplier of the Gauss term in the RHS of (\ref{DG}) is  $(\even) r^{-1} +O(r^{-1-\e})$ and hence satisfies the conditions of section \ref{gausssec}.  We now show that the surface term in (\ref{DG}) vanishes. %We compute separately the surface terms arising form $\L_{\vec{N}} \Lambda$ and  $[\LN, \Lambda]$. Since each of them are $(\even) r^{-1}$,  only the zeroth order part of the electric field $E^a_i$ in (\ref{DG}) contributes. 
 The first term can be written as:
\be
\oint_{\infty} d S_a E^a_i \L_{\vec{N}} \Lambda^i = - \oint_{\infty} d S_a \L_{\vec{R}} \Eo^a_i \Lambda^i ,
\ee
where we used the fact that $\L_{\vec{N}} \Lambda = (\even) r^{-1}+O(r^{-1-\e})$ and  Eq. (\ref{intSlieN}). For the second term we have,
\ba
-\oint_{\infty} d S_a E^a_i [\LN, \Lambda]^i &= &-\oint_{\infty} d S_a \epsilon_{ijk} \Eo^a_i \LN^j \Lambda^k \\
& = &-\frac{1}{2}\oint_{\infty} d S_a \epsilon_{ijk} \e_{jmn} \Eo^a_i \Eo_b^m \L_{\vec{R}} \Eo^b_n \Lambda^k \\
& = & \frac{1}{2} \oint_{\infty} dS_a( \L_{\vec{R}} \Eo^a_i \Lambda^i - \Eo_b^i \L_{\vec{R}} \Eo^b_j \Eo^a_j \Lambda^i)\\
&=& \oint_{\infty} d S_a \L_{\vec{R}} \Eo^a_i \Lambda^i ,
\ea
where we used the fact that $[\LN, \Lambda]=(\even) r^{-1} + O(r^{-1-\e})$ and   $\L_{\vec{R}} \Eo^b_j \Eo^a_j = - \Eo^b_j \L_{\vec{R}}  \Eo^a_j$. The two terms cancel each other and we conclude that $D[\vec{N}]$ weakly commutes with the Gauss constraint.

\subsubsection{Boosts}
Expressing the boundary term in (\ref{H1}) as a volume integral, one can isolate the divergent term:
\be
H_1[\N]= -\frac{1}{8 \pi G} \int_{\Sigma}  \Do_a \N \e_{i j k}A^i_b E^{a}_j E^{b}_k + \ldots  \label{Cprime}
\ee
where the dots represents  terms whose integral is convergent for  $\N \to \beta_I x^I$. The divergent piece can be removed  by subtracting a Gauss term $G[\Lambda_B]$ with 
\be
\LB^i= \LBo^i + \Lambda^i, \label{defLB}
\ee
where
\be
\LBo^i := \beta_I \Eo^I_i = \Do_a B \Eo^a_i \label{defLBo}
\ee
is a constant,  zeroth order term and  $\Lambda^i$  a `pure gauge' multiplier as in (\ref{fallofflambda}), (\ref{paritylambda}).
As in the case of rotations,  the Gauss term introduces a divergent and non-differentiable piece that can be removed by an appropriate surface term. The final expression is: %\footnote{As in the case of rotations  (\ref{finalD}), the expression ``$H_1[\N] -\gamma G[\LB]$'' is a shorthand notation for a single integral, see footnote \ref{fnote2}.}
\be
H[\N]:=H_1[\N] -\gamma G[\LB]+ \frac{1}{8 \pi G }\oint_{\infty} d S_a  E^a_i \LBo^i .\label{finalC}
\ee
We now verify the expression  satisfies (i) and (ii).  In section \ref{sec4} we  show (\ref{finalC}) agrees, modulo pure gauge Gauss constraint terms, with ADM and Thiemann's expressions.  

As we did for the rotations,  let us express  (\ref{finalC}) as a volume integral $H[\N]= (8 \pi G)^{-1}\int_\Sigma \rho$ with:
\be
\rho =  - N \e_{ijk} A_b \Do_a(E^a_j E^b_k) - \Do_a \N \e_{i j k}A^i_b E^{a}_j E^{b}_k  - \e_{ijk} \LB^i  A^j_b E^b_k-(\LB^i-\LBo^i) \partial_a E^i + \ldots . \label{rho}
\ee
The dots indicate the   `NAAEE' term coming from the non-abelian part of $F_{ab}$ and the `Lorentzian'  `NKKEE' term.  The former is manifestly finite and differentiable. The latter is also finite and differentiable by the discussion given in the beginning of \ref{sec3C}. We then focus on the terms displayed in (\ref{rho}). The first and last terms are easily verified to give a finite and differentiable contribution to $H$.  By construction, the potentially  divergent contributions from the second and third term cancel out, as can be verified by using (\ref{lP}) and (\ref{defLBo}).  We conclude that $H[\N]$ is finite and differentiable. One can further verify that all possible contributions to the Hamiltonian vector field are such that they preserve the fall-off and parity conditions so that $H[\N]$ satisfies (ii).  

We finally note that if $B=0$, then $\LBo^i=0$ and we recover the generator of time translations (\ref{H1}), up to a possible `pure gauge' Gauss term.

\section{Comparison with ADM and Thiemann's expressions} \label{sec4}
\subsection{Diffeomorphism constraint, asymptotic translations and rotations}
We give a quick re-derivation of Thiemann's expressions based on the ADM ones, and show they coincide (up to pure gauge Gauss terms) with the expressions from last section.  
We start with the ADM  generator (\ref{Dtotal}):
\be
 \Dadm[\vec{N}] = \frac{1}{16 \pi G} \int_{\Sigma} \pi^{ab}\L_{\vec{N}} q_{ab} ,\label{admD}
\ee
with a general shift of the form (\ref{asymshift}),  and seek to rewrite it in terms of   $(A_a^i,E^a_i)$ variables.  A small computation shows the integrand in (\ref{admD}) can be rewritten as, 
\be
\pi^{ab}\L_{\vec{N}} q_{ab} = -2 q^{-1/2} K_{ab} \L_{\vec{N}} q q^{ab}, \label{integrandDadm}
\ee
where we used $\pi^{ab}=q^{1/2}(K^{ab}-Kq^{ab})$. Performing the substitution,
\ba
q q^{ab} & = & E^a_i E^b_i ,\label{qEE}\\
q^{-1/2} K_{ab} & = & E^j_{(a}K^j_{b)},
\ea
in (\ref{integrandDadm}), the ADM generator (\ref{admD}) becomes:
\be
\Dadm[\vec{N}]= -\frac{1}{8 \pi G} \int_{\Sigma}( K_a^i \L_{\vec{N}} E^a_i + \e_{i j k} \LT^i K_a^j E^a_k),
\ee
where we have defined
\be
\LT^i := \frac{1}{2} \e_{ijk}E^j_a \L_{\vec{N}} E^a_k. \label{lambdaT}
\ee
Finally, substituting $K_a^i=\gamma^{-1}(A_a^i-\Gamma_a^i)$ we recover the  expression given in \cite{Taf}:
\be
\Dadm[\vec{N}]= D_1[\vec{N}] - G(\LT) +\frac{1}{8 \pi G \gamma} \int_{\Sigma} \Gamma_a^i \L_{\vec{N}} E^a_i ,  \label{Dadm}
\ee
with $D_1$ and $G$ given by  (\ref{D1}) and (\ref{gauss}) respectively. As we shall see, the last term in (\ref{Dadm}) can be written as a surface term.  In appendix \ref{tteqadm} we display this surface term in the form  given in \cite{Taf}.  

In the following  we discuss  asymptotic rotations and translations  separately. We will write the last term in  (\ref{Dadm})  in a way that will facilitate the comparison with the expressions of section \ref{sec3}.

%For the purposes of comparing (\ref{Dadm}) with the expressions of section \ref{sec3},  a different  form of the surface term is  convenient. This is described in the following two subsections, where  asymptotic rotations and asymptotic translations are treated separately.

%For the comparison with the expressions of section \ref{sec3}   we will use a different form of expressing this surface term. In particular we will discuss separately the cases of asymptotic rotations and translations. 

%consider the case of asymptotic rotations to compare (\ref{Dadm}) with the generator (\ref{finalD}). For completeness we will also verify agreement of (\ref{Dadm}) with (\ref{D1}) for the case of asymptotic translations. 

% In the following  We now focus attention in the case of asymptotic rotations, where we will show that the generator given in the previous section (\ref{finalD}) coincides, modulo a Gauss term, with the ADM one.  For completeness we will also  re-derive the result that, for asymptotic translations, the generator (\ref{D1}) coincides, modulo a Gauss term, with (\ref{Dadm}).

\subsubsection{Rotations}
Consider the case where $\alpha^I=0$ so that ,
\be
N^I = R^I + S^I+O(r^{-\e}) , \label{asymrot}
\ee
with $R^I$ as in (\ref{Rbeta}) and $S^I$ as in (\ref{oddst}). The last term in the RHS of  (\ref{Dadm}) can be shown to be a pure  boundary term as follows. First integrate by parts,
\be
\int \Gamma_a^i \L_{\vec{N}} E^a_i = \oint dS_a N^a \Gamma_b^i E^b_i - \int E^a_i \L_{\vec{N}} \Gamma_a^i  . \label{intbyparts}
\ee
The surface integral vanishes since  $dS_a N^a=(\even)+O(r^{-\e})$ ,  $\Gamma_b^i=(\odd) r^{-2}$ and $E^b_i$ as in (\ref{falloffE}).  For the second integral we use the identity (see appendix \ref{EdGapp}),
\be
E^a_i \delta \Gamma_a^i = -\frac{1}{2}\e_{ijk}\partial_a ( E^a_i E^j_b \delta E^b_k) ,\label{EdG}
\ee
to write it as a boundary term:
\ba
- \int E^a_i \L_{\vec{N}} \Gamma_a^i  &=& \frac{1}{2} \oint_{\infty} dS_a \e_{ijk}  E^a_i E^j_b \L_{\vec{N}} E^b_k \\
& =& \oint_{\infty} dS_a E^a_i \LT^i , \label{intbyparts2}
\ea
where in the second equality we used the definition of $\LT$ (\ref{lambdaT}). We thus obtain:
\be
\Dadm[\vec{N}]=D_1[\vec{N}] - G(\LT) +\frac{1}{8 \pi G \gamma} \oint_{\infty} dS_a E^a_i \LT^i  \label{Daf}.
\ee
Expression (\ref{Daf}) resembles that of the generator (\ref{finalD}) given in the previous section. Choosing for simplicity $\LN=\LNo$ in (\ref{finalD}), the difference between the two is:
\be
 D[\vec{N}] - \Dadm[\vec{N}] = G[\lN] - \frac{1}{8 \pi G \gamma} \oint_{\infty} dS_a E^a_i \lN^i \label{F} 
\ee
with
\be
\lN^i := \LT^i - \LNo^i =(\even) r^{-1} +O(r^{-1-\e}).
\ee
In appendix \ref{bdyzeroapp} we show that
\be
\oint_{\infty} dS_a E^a_i \lN^i =0 \label{bdyzero}
\ee
and hence the difference (\ref{F}) is a  `pure gauge' Gauss term $G(\lN)$ (with phase space dependent multiplier).

\subsubsection{Translations}
For completeness we re-derive the result that for asymptotic translations the generator (\ref{D1}) coincides, modulo a Gauss term, with the ADM generator (\ref{Dadm}). Since the  considerations from the previous section already account for the $S^I$ and $O(r^{-\e})$ terms in the shift (see appendix \ref{bdyzeroapp}), we now restrict attention to shifts of the form
\be
N^I = \alpha^I + O(r^{-1-\e}). \label{shifttrans}
\ee
The surface term that was dropped in (\ref{intbyparts}) no longer vanishes and so the last term in the RHS of  (\ref{Dadm}) now becomes
\be
 \int_{\Sigma} \Gamma_a^i \L_{\vec{N}} E^a_i =\oint dS_a (N^a \Gamma_b^i E^b_i  +E^{a}_i \LT^i). \label{translbdy}
\ee
For the first term in (\ref{translbdy}) we write:
\ba
\Gamma_b^i E^b_i =-\frac{1}{2} \e_{ijk} E^b_i E^j_c  \Do_b E^c_k \label{GE}
\ea
(this follows from Eq. (\ref{spinc}) by noting that the additional $(D_b-\Do_b)$ contribution vanishes). For the second term we write
\be
\LT^i = \frac{1}{2}\e_{ijk} E_c^j N^b \Do_b E^c_k +O(r^{-2-\e}) \label{tbdy2}
\ee
where we used that $\Do_a N^b =O(r^{-2-\e})$ for the shift (\ref{shifttrans}) .   Since $\Do_a E^b_i=(\odd) r^{-2}$, the triads in (\ref{GE}) and (\ref{tbdy2}) that are not being derivated can be set to their zeroth order value.   Defining
\be
X^a:= \frac{1}{2}\epsilon_{ijk}\Eo^a_i \Eo^j_c E^c_k,
\ee
the surface integral (\ref{translbdy}) becomes
\be
 \oint dS_a (-N^a \Do_b X^b + N^b \Do_b X^a) =  \oint dS_a (-N^a \Do_b X^b +\L_{\vec{N}} X^a) =0 
\ee
where we used Eq. (\ref{intSlieN}). Thus, the surface term vanishes and  the generators (\ref{Dadm}) and (\ref{D1}) differ by a pure gauge Gauss term.

\subsection{Hamiltonian constraint, asymptotic time translations and boosts}
We start with a quick re-derivation of the Hamiltonian in Ashtekar-Barbero variables in order to ensure no further subtleties arise from the `KKEE' term. Let 
\be
\Hadm[\Nadm] =\Hadmo[\Nadm] +S[\Nadm] \label{Hadm}
\ee
be the  full ADM Hamiltonian (\ref{Htotal}) with $\Hadmo[\Nadm]$  given by Eq. (\ref{H0}) and  $S[\Nadm]$  the surface term in (\ref{Htotal}). The lapse is taken to be $\Nadm=q^{1/2} N$ with $N$ the density weight -1 lapse  satisfying (\ref{lP}).

For the integrand of $\Hadmo$ we use  the identities (see for instance \cite{Tbook}):
\ba
- q R & = & \e_{i j k} F_{ab}^i E^a_j E^b_k + 2  D_a (E^a_i \D_b E^b_i)  -2 \gamma^2 K^i_{[a}K^j_{b]}E^a_i E^b_j  \label{qR} \\
\pi_{ab}\pi^{ab}-\tfrac{1}{2}\pi^2 & =& -2  K^i_{[a}K^j_{b]}E^a_i E^b_j +\frac{1}{2\gamma^{2}}\D_a E^a_i \D_b E^b_i , \label{pi2}
\ea
where   $D_a$ is the derivative compatible with $q_{ab}$ (\ref{qEE}). For the surface term $S[\Nadm]$ we use the result derived in \cite{Taf}:
\be
S[q^{1/2} N] = \frac{1}{8 \pi G}\oint_{\infty} d S_a (- N E^a_i \partial_b E^b_i + D_b N E^b_i (E^a_i -\Eo^a_i)). \label{Sadm}
\ee
We now use (\ref{qR}), (\ref{pi2}), (\ref{Sadm}) to express  (\ref{Hadm}) in $(A_a^i,E^a_i)$ variables. Subtracting for convenience the `pure gauge' Gauss piece arising from the second term in (\ref{pi2}) and integrating by parts the middle  term in (\ref{qR}), one obtains:
\ba
\Hadm'[q^{1/2} N] & := &\Hadm[q^{1/2} N]- ({4 \gamma})^{-1}G[N \D_a E^a]  \\
&=& H_1[\N] -\gamma G[\Lambda_N] +\frac{1}{8 \pi G}\oint_{\infty} d S_a \Lambda_N^i (E^a_i -\Eo^a_i) ,\label{CT}
\ea
with  $H_1$ given by (\ref{H1}), $G$ as in (\ref{gauss})  and 
\be
\Lambda_N^i:=D_a N E^a_i.
\ee
Expression (\ref{CT}) corresponds to the one given in \cite{Taf}, written in a way that will facilitate comparison with $H[N]$ (\ref{finalC}).  Subtracting $0=\oint_{\infty} d S_a \LBo^i \Eo^a_i$ in $H[N]$ so that the surface term in (\ref{finalC}) involves the difference $(E^a_i - \Eo^a_i)$ as in (\ref{CT}), we find:
\be
\Hadm'[q^{1/2} N]-H[N]= -\gamma G[\Lambda_N-\LB]+ \frac{1}{8 \pi G }\oint_{\infty} d S_a (\Lambda_N^i-\LBo^i) (E^a_i -\Eo^a_i). \label{diffC}
\ee
The difference $\Lambda^i_N-\LB^i$ can be seen to be $(\even) r^{-1}$ as follows. First write
\be
\Lambda_N^i = \Do_a N E^a_i + q^{-1/2} \Do_a q^{1/2} N E^a_i = \Do_a N E^a_i +(\even) r^{-1} \\
\ee
since $\Do_a q^{1/2} = (\odd)r^{-2}$ and $N=(\odd) r$. Finally, 
\ba
\Do_a N E^a_i-\LBo^i &= &  \Do_a N E^a_i- \Do_a B \Eo^a_i \\
&=& \Do_a N (E^a_i - \Eo^a_i) + \Do_a(N-B) \Eo^a_i \\
&=& (\even) r^{-1}.
\ea
It then follows that the surface term in (\ref{diffC}) vanishes and  the difference (\ref{diffC}) is given by the `pure gauge' Gauss term  $-\gamma G[\Lambda_N-\LB]$. Finally, we note that even though we were here mostly interested in the case of boosts, the lapse $N$ above is of the general form (\ref{lP}).  If  $B=0$  the expressions reduce (modulo a pure gauge Gauss term) to the generator of asymptotic time translations $H_1[N]$ (\ref{H1}). \\

\textbf{Acknowledgements:} 
I am grateful to  Madhavan Varadarajan for his generous  guidance over this project, and to Alok Laddha and Sandipan Sengupta for helpful discussions. I would also like to thank AL, MV and Casey Tomlin  for their comments on the manuscript.

\appendix
\section{Assorted results}

\subsection{Eq. \ref{EdG}} \label{EdGapp}
We starting from an identity given in \cite{Taf,Tbook}: 
\be
E^a_i \delta \Gamma_a^i   =  \frac{1}{2}\partial_a (\eta^{abc}e^j_b \delta e^j_c) \label{EdGT}, 
\ee
and rewrite the expression inside the derivative as:
\ba
\eta^{abc}e^j_b \delta e^j_c & = & e^j_b \delta ( \eta^{abc} e^j_c ) \\
& = & q^{-1/2} e^j_b \delta ( q^{1/2} \eta^{abc} e^j_c) \\
& = & E^j_b \delta (-\e_{ijk}E^a_i E^b_k)
\ea
using the last expression back in (\ref{EdGT}) we obtain (\ref{EdG}).

\subsection{Integration by parts  formulas}
Let  $\rho$ be a density one scalar and  $X^a$  a density one vector field. $\rho$ is dual to the 3-form $\rho \eta_{abc}$ and $X^a$ dual to the 2-form $\omega_{ab}  :=  \eta_{abc} X^c $ so that $X^a  =  \frac{1}{2}\eta^{abc} \omega_{bc}$.

Stokes theorem for the integral of $\rho= \partial_a X^a$ over a volume $V$ reads
\be
\int_V \partial_a X^a = \int_{\partial V} dS_a X^a.
\ee
In particular, if $\vec{N}$  is a vector field then $\L_{\vec{N}} \rho = \partial_a (\rho N^a)$ and one has:
\be
\int_V \L_{\vec{N}} \rho = \int_{\partial V} dS_a N^a \rho.
\ee
The Lie derivative of $X^a$ along a vector field $\vec{N}$ can be written as
\be
\L_{\vec{N}} X^a = N^a \partial_b X^b + 2 \partial_b( X^{[a} N^{b]}), \label{lieX}
\ee
where the second term is a total derivative: $2 \partial_b( X^{[a} N^{b]})=\eta^{abc}\partial_b Y_c$ with $Y_c = \eta_{cde} X^d N^e$.  Integrating (\ref{lieX}) over a two-surface $S$ without boundary we obtain the relation:
\be
\oint_S dS_a \L_{\vec{N}} X^a =  \oint_S dS_a N^a \partial_b X^b. \label{intSlieN}
\ee

\subsection{Surface term of Eq. (\ref{Dadm}) in terms of $\Gamma_a^i$'s} \label{tteqadm}
For a general asymptotic shift, $N^I = \alpha^I+ R^I + S^I+O(r^{-\e})$, 
Eqns. (\ref{intbyparts}) and (\ref{intbyparts2}) expresses  the last term in (\ref{Dadm}) as a pure surface term:
\be
 \int_{\Sigma} \Gamma_a^i \L_{\vec{N}} E^a_i =\oint dS_a (N^a \Gamma_b^i E^b_i  +E^{a}_i \LT^i) .\label{translbdy2}
\ee
If we write $\L_{\vec{N}} E^a_k$ in (\ref{lambdaT}) in terms of the derivative $D_a$ compatible with $q_{ab}$ (\ref{qEE}), the second term on the RHS of (\ref{translbdy2}) is given by:  
\be
\oint_{\infty} dS_a E^a_i \LT^i = \frac{1}{2} \oint_{\infty} d S_a\e_{ijk}E^a_i E^j_c N^b D_b E^c_k - \frac{1}{2} \oint_{\infty}dS_a \e_{ijk}E^a_i E^j_c E^b_k D_b N^c. \label{twoint}
\ee
The integrand of the second term in (\ref{twoint}) can be written as the total derivative $\eta^{a b c} D_b N_c$ with $N_c:=q_{cd}N^d$ and hence the integral vanishes.  The first term can be cast in terms of the spin connection by use of the formula
\be
\Gamma_b^i= -\frac{1}{2} \e_{ijk} E^j_c D_b E^c_k. \label{spinc}
\ee
Doing so one obtains:
\be
\oint_{\infty} dS_a E^a_i \LT^i  = -\oint_{\infty} dS_a \Gamma^i_b N^b E^a_i , \label{bdymett}
\ee
which together with the first term in the RHS of (\ref{translbdy2}) correspond to form of the surface term  given in \cite{Taf}.

\section{Eq. (\ref{bdyzero})} \label{bdyzeroapp}
Let us denote the `pure gauge' part of the  shift (\ref{asymrot}) by $\vec{\nu}$ so that:
\be
N^I= R^I+\nu^I \; ;  \quad \nu^I = S^I +O(r^{-\e}).
\ee
Let
\ba
\lR^i &: = &\frac{1}{2} \e_{ijk}(E^j_b \L_{\vec{R}} E^b_k- \Eo^j_b \L_{\vec{R}} \Eo^b_k) ,\\
\Lnu^i &:=& \frac{1}{2} \e_{ijk}E^j_b \L_{\vec{\nu}} E^b_k  ,\label{lnu}
\ea
so that
\be
\lN^i = \lR^i +\Lnu^i . \label{lam}
\ee
We now show that the contribution to the surface integral (\ref{bdyzero}) from each term in (\ref{lam}) vanishes. The contribution coming from $\Lnu^i$ can be written as in (\ref{twoint}) with $\vec{N}=\vec{\nu}$. The second term on the RHS of  (\ref{twoint}) is again a total derivative whose integral vanishes, whereas the integrand of the first term is now $(\odd) r^{-2}+ O(r^{-2-\e})$ and hence also vanishes. 

To study the contribution from $\lR^i$, let us write the triad as 
\be
  E^a_i = \Eo^{a}_i +\Ei^a_i+ \Eii^a_i ,
\ee
where  $\Ei^a_i$ represents the  $(\even) r^{-1}$ term in (\ref{falloffE}) and $\Eii^a_i$ the remaining  $O(r^{-1-\e})$ piece.
Let
\be
  E^i_a = \Eo^i_a +\Ei^i_a+ \Eii^i_a 
\ee
denote the analgous  expansion of the inverse triad so that
\ba
\Ei^i_a &= & - \Eo^i_b \Eo_a^j \Ei^b_j  \\ 
\Eii^i_a &= & - \Eo^i_b \Eo_a^j \Eii^b_j+\Eo^i_b \Eo_a^j \Eo^k_c \Ei^c_j \Ei^b_k +O(r^{-2-\e}).\label{Einv2}
\ea
The corresponding expansion for $\lR$ is:
\be
\lR^i = \frac{1}{2}\e_{ijk}\left( \Eo_a^j \L_{\vec{R}} \Ei^a_k + \Ei_a^j \L_{\vec{R}} \Eo^a_k + \Ei_a^j \L_{\vec{R}} \Ei^a_k + \Eo_a^j \L_{\vec{R}} \Eii^a_k + \Eii_a^j \L_{\vec{R}} \Eo^a_k\right) + O(r^{-2-\e}). \label{lRexp}
\ee
Parity conditions imply that  the nontrivial contributions to the integral come from the last two terms in (\ref{lRexp}):
\be
\oint dS_a E^a_i \lR^i = \frac{1}{2} \oint dS_a \e_{ijk}\Eo^a_i ( \Eo_b^j \L_{\vec{R}} \Eii^b_k + \Eii_b^j \L_{\vec{R}} \Eo^b_k). \label{intlr}
\ee
We now integrate by parts the first term (using  Eq. (\ref{intSlieN}) and $dS_a R^a =0$) and use (\ref{Einv2}) for $\Eii_b^j$ in the second term (only the first term in (\ref{Einv2}) contributes, since the  `$\Ei \Ei$' piece is even) to get:
\be
\oint dS_a E^a_i \lR^i = -\frac{1}{2} \oint dS_a \e_{ijk}(  \L_{\vec{R}}(\Eo^a_i \Eo_b^j)  \Eii^b_k +\Eo^a_i \Eo^j_c \Eo_b^l \Eii^c_l \L_{\vec{R}} \Eo^b_k). \label{intlr2}
\ee
Using
\ba
\L_{\vec{R}}(\Eo^a_i \Eo_b^j) &=& -\Eo^c_i \Eo_b^j \Do_c R^a +\Eo^a_i \Eo_c^j \Do_b R^c \\
\L_{\vec{R}} \Eo^b_k &=& - \Eo^d_k \Do_d R^b,
\ea
(\ref{intlr2}) can be written as (after renaming some indices and using $\qo_{ab}$ to raise and lower some indices):
\be
\oint dS_a E^a_i \lR^i = \frac{1}{2} \oint dS_a B^{abk} \qo_{be} \Eii^e_k
\ee
with
\be
B^{ab k}= \e_{ijk}\Eo^i_c \Eo^b_j \Do^c R^a - \e_{ijk}\Eo^{a}_i \Eo^j_c \Do^b R^c+ \e_{ijl} \Eo^a_i \Eo^{b}_l \Eo^j_c \Eo^k_d \Do^{d}R^c. \label{Babk}
\ee
By writing (\ref{Babk}) in cartesian coordinates so that $\Eo^a_i = \delta^a_i$, etc. and using that  $\Do^a R^b = \epso^{abc}\phi_c$ for some constant $\phi_c$, one finds that (\ref{Babk}) identically vanishes:
\ba
B^{ab k} & = & \e_{cb k} \e_{ca d} \phi_d- \e_{ack}\e_{bcd} \phi_d +\e_{acb} \e_{kcd} \phi_d \\
& = &[ (\delta_{ab}\delta_{kd}-\delta_{ak}\delta_{bd}) -(\delta_{ab}\delta_{kd} -\delta_{ad} \delta_{kb})+ (\delta_{ak}\delta_{bd}-\delta_{ad}\delta_{kb}) ] \phi_d =0.
\ea
This concludes the proof of Eq. (\ref{bdyzero}).

\end{document}